\documentclass[aps,a4paper,pra,floats,amsfonts]{revtex4}

\usepackage{amsmath}
\usepackage{amsfonts}
\usepackage{amssymb}

\newcommand{\be}{\begin{equation}}
\newcommand{\ee}{\end{equation}}
\newcommand{\bea}{\begin{eqnarray}}
\newcommand{\eea}{\end{eqnarray}}
\newcommand{\bean}{\begin{eqnarray*}}
\newcommand{\eean}{\end{eqnarray*}}

\begin{document}

  \title{Stronger Violation of Local Theories with Equalities}
  \author{Hossein Movahhedian}
  \affiliation{Department of Physics, Shahrood University of Technology,
    Seventh Tir Square, Shahrood, Iran}
  \email{hossein_movahhedian@catholic.org}

  \begin{abstract}
    Bell type inequalities are used to test local realism against quantum
    theory. In this paper, we consider a two party system with two settings and
    two possible outcomes on each side, and derive {\em equalities} in local
    theories which are violated by quantum theory by a factor of 1.522
    tolerating $0.586$ fraction of white noise admixture which is twice that of
    the previous results.
    \begin{center} \today \end{center}
  \end{abstract}

\maketitle

\section{Introduction \label{sec_01}}
  The idea of local realism opened its way into quantum theory by the work of
  Einstein, Podolsky and Rosen which is well known as EPR paradox
  \cite{EPR_35}, though it is not really a paradox, because it has been argued
  recently that the assumptions of EPR were wrong~\cite{Peres_A_03}. However it
  was left to John S. Bell who derived an inequality based on local theories
  and proved that it was violated by statistical predictions of quantum theory
  \cite{BEL_J_S_64,BEL_J_S_66}. Since then many attempts have been made to
  derive Bell type inequalities which are violated by a stronger factor, so
  that it can be tested in the real experiments in which errors are inevitable.
  Among these are Clauser, Horne, Shimony and Holt inequality, the so called
  CHSH inequality \cite{CHSH_69,CH_74}. 

  Nowadays, with the growing power of computers, the numerical methods have
  attracted attentions for constructing these inequalities as much as possible
  \cite{NOC_C_95,MOV_H_98,KGZMZ_00} though analytical approaches are in
  progress too \cite{CGLMP_02}; with the hope that some of which could be
  violated by a stronger factor.

  In this paper we first introduce a simple but general method for constructing
  Bell expressions which can be applied to two party experiments each measuring
  two observables with different outputs. Generalization to more parties and/or
  more observables is straightforward. Then in a special case of a two party
  system with two settings and two possible outcomes on each side, we use these
  expressions and show that according to local theories there exist
  {\em equalities} which are violated by quantum theory by a stronger factor
  than Bell type inequalities. Meanwhile we derive an inequality in this case
  which is violated by a factor of 1.621 tolerating $0.293$ fraction of white
  noise admixture. Although in this case the tolerance is the same as those
  derived previously in the literature the range of violation is greater.

\section{Derivation of Bell Type Expressions \label{sec_02}}
  Let's consider a two party system with no known interaction between its
  parts. Suppose the left party, say $A$ performs two possible measurements $a$
  with outcomes \mbox{$i \in \{0, \cdots, m - 1$\}} and $a'$ with outcomes
  \mbox{$i' \in \{0, \cdots, m' - 1$\}}. Similarly the right party, say $B$,
  performs two possible measurements $b$ with outcomes \mbox{$j \in \{0,
  \cdots, n - 1$\}} and $b'$ with outcomes
  \mbox{$j' \in \{0, \cdots, n' - 1$\}}.
  For simplicity, from now on we use unprimed/primed variables and indexes for
  the first/second measurement for each party whenever applicable and label
  such a system as $mn\!\otimes\!m'n'$. As a consequence of locality, for
  measurements which are not simultaneous, there exists a probability
  $q_{aa'bb'}^{ii'jj'}$ defined as the probability that measurements of $a$
  results $i$, $a'$ results $i'$, $b$ results $j$ and $b'$ results $j'$. The
  total number of $q$'s, $N_Q$, is
  \mbox{$N_Q=m\!\times\! m'\!\times\! n\!\times\! n'$}. As $q$'s are all
  exclusive and cover all probable events, with the assumption that the
  probability distribution for each measurement is normalized we have
  \be
    \sum_{i,i',j,j'} q_{aa'bb'}^{ii'jj'} = 1.
    \label{eq_001}
  \ee
  These $q$'s are the building blocks of all possible Bell expressions for
  local theories.

  If $P_{ab}^{i j}$ denotes the probability that in a particular experiment,
  $A$ measures $a$ with outcome $i$ and $B$ measures $b$ with outcome $j$, we
  can write
  \be
    P_{ab}^{ij} = \sum_{i',j'} q_{aa'bb'}^{ii'jj'},
       \hspace{1em} \text{and so on ...}
    \label{eq_002}
  \ee

  There are totally $N_P$ number of $P$'s, where \mbox{$N_P=(m+m')(n+n')$}.
  Eq.~(\ref{eq_002}) in matrix form would become
  \be
    \mathbf{P}=\mathbf{MQ}
    \label{eq_003}
  \ee
  where $\mathbf{P}$ is an $N_P\!\times\! 1$ column matrix, $\mathbf{Q}$ is an
  $N_Q\!\times\! 1$ column matrix and $\mathbf{M}$ is the conversion matrix
  with dimension $N_P\!\times\! N_Q$. These $P$'s are not all independent. The
  rank of the matrix $\mathbf{M}$, denoted by $N_I$, is the number of
  independent $P$'s. For
  $22\!\otimes\!22$, $23\!\otimes\!22$ and $23\!\otimes\!23$
  the rank of the  conversion matrix, $\mathbf{M}$, is 9, 12 and 16
  respectively. Please note that these do not agree with the results obtained
  in \cite{CO_D_GI_N_03} in which the number of independent $P$'s for
  $22\!\otimes\!22$ case is predicted to be 8, and for $23\!\otimes\!22$ case
  it is 11 (for $m=2$, $m'=3$, $n=2$ and $n'=2$) or 14
  (for $m=2$, $m'=2$, $n=2$ and $n'=3$) and for $23\!\otimes\!23$ case it is 20
  (for $m=2$, $m'=3$, $n=2$ and $n'=3$) or 19 (for $m=3$, $m'=2$, $n=2$ and
  $n'=3$).

  To prove these numerical results analytically we search for constraints which
  justify the rank of the matrix $\mathbf{M}$. Two groups of constrains are
  directly derived from eqs.~(\ref{eq_001}) and (\ref{eq_002}).

  The first is {\em normalization} of probability distribution in a
  measurement. That is
  \bea
    \sum_{i,j} P_{ab}^{ij} & = & 1,
    \label{eq_004} \\
    \sum_{i,j'} P_{ab'}^{ij'} & = & 1,
    \label{eq_005} \\
    \sum_{i',j} P_{a'b}^{i'j} & = & 1,
    \label{eq_006} \\
    \sum_{i',j'} P_{a'b'}^{i'j'} & = & 1.
    \label{eq_007}
  \eea
  However, with the assumption of CH (see \cite{CH_74}) we can write
  \be
    P_{ab}^{ij} P_{a'b'}^{i'j'} = P_{a'b}^{i'j} P_{ab'}^{ij'} =
    q_{aa'bb'}^{ii'jj'},
    \label{eq_008}
  \ee
  and from eq.~(\ref{eq_001}) we have
  \be
    \sum_{i,i',j,j'} P_{ab}^{ij} P_{a'b'}^{i'j'} =
    \sum_{i,i',j,j'} q_{aa'bb'}^{ii'jj'} = 1,
    \label{eq_009}
  \ee
  \be
    \sum_{i,i',j,j'} P_{a'b}^{i'j} P_{ab'}^{ij'} =
    \sum_{i,i',j,j'} q_{aa'bb'}^{ii'jj'} = 1.
    \label{eq_010}
  \ee
  So, for instance, eqs.~(\ref{eq_004}), (\ref{eq_005}) and (\ref{eq_009}) can
  be considered as independent and these equations impose only 3 independent
  constraints on $P$'s. Please note that in this case eq.~(\ref{eq_010}) is
  automatically satisfied and eq.~(\ref{eq_006}) can be derived from
  eqs.~(\ref{eq_005}) and (\ref{eq_010}). Similarly eq.~(\ref{eq_007}) can
  be derived from eqs.~(\ref{eq_004}) and (\ref{eq_009}).

  The second group of constraints derived from eqs.~(\ref{eq_001}) and
  (\ref{eq_002}) are
  \bea
    \sum_{j} P_{ab}^{ij} & = & \sum_{j'} P_{ab'}^{ij'},
       \label{eq_011} \\
    \sum_{j} P_{a'b}^{i'j} & = & \sum_{j'} P_{a'b'}^{i'j'},
       \label{eq_012} \\
    \sum_{i} P_{ab}^{ij} & = & \sum_{i'} P_{a'b}^{i'j},
       \label{eq_013} \\
    \sum_{i} P_{ab'}^{ij'} & = & \sum_{i'} P_{a'b'}^{i'j'},
       \label{eq_014}
  \eea
  which imply {\em no-signaling}.
  The total number of constraints from the above equations add up to
  \mbox{$m+m'+n+n'$}. However in each group of the above constraints, one of
  them is a linear combination of the other, for example from
  eqs.~(\ref{eq_004}) and~(\ref{eq_005}) one can write
  \be
    \sum_{i,j} P_{ab}^{ij} = \sum_{i,j'} P_{ab'}^{ij'},
       \label{eq_015}
  \ee
  and from eq.~(\ref{eq_011}) we have
  \be
    \sum_{i\neq l, j} P_{ab}^{ij} =
    \sum_{i\neq l,j'} P_{ab'}^{ij'}.
       \label{eq_016}
  \ee
  Subtracting eq.~(\ref{eq_016}) from eq.~(\ref{eq_015}) results in
  \be
    \sum_{j} P_{ab}^{lj} =
    \sum_{j'} P_{ab'}^{lj'}.
       \label{eq_017}
  \ee
  which is one of the constraint in eq.~(\ref{eq_011}). So the constraints from
  no-signaling would be \mbox{$(m-1)+(m'-1)+(n-1)+(n'-1)$}.

  The total number of constraints, $N_C$, given by
  eqs.~(\ref{eq_004} -- \ref{eq_014}) would become
  \be
    N_C = m + m' + n + n' - 1.
       \label{eq_018}
  \ee
  So there are only $N_I$ number of independent $P$'s, where
  \bea
    \hspace*{-3 ex}
    N_I & \!\! = & \!\! N_P - N_C \nonumber \\
        & \!\! = & \!\! (m+m')(n+n') - (m+m'+n+n' - 1).
       \label{eq_019}
  \eea
  which agrees with the numerical results mentioned before and is symmetric
  with respect to interchanging $m$ and $m'$ (and of course $n$ and $n'$).
  Clearly without CH assumption {\it i.e.} eq.~(\ref{eq_008}) we couldn't get
  this result. We would like to emphasize that this assumption is only used
  here to prove eq.~(\ref{eq_019}). It has nothing to do with the rest
  of this paper, especially the main results which will be obtained later.

  Generally if ${\mathbb B}$ is a Bell expression for local theories and
  $-d \le {\mathbb B} \le c$
  with non-negative $c$ and $d$, then $\mathbb{B}$ must satisfy
  \bea
    {\mathbb B} & = & \sum_{s,t,k,l} \lambda_{stkl}P_{st}^{kl} \nonumber \\
             & = & \sum_{i,i',j,j'} (\mu_{ii'jj'} - \nu_{ii'jj'})
                   q_{aa'bb'}^{ii'jj'},
             \hspace{15pt} \mu_{ii'jj'} \neq \nu_{ii'jj'}
    \label{eq_020}
  \eea
  where $c(d)$ is the greatest of non-negative real numbers $\mu$'s($\nu$'s).

  We have solved eq.~(\ref{eq_003}) numerically for $22\!\otimes\!22$,
  $23\!\otimes\!22$ and $23\!\otimes\!23$ cases to find all possible
  expressions that satisfy eq.~(\ref{eq_020}) and the complement of each one,
  {\it i.e.} all pair of expressions whose sum add up to 1. However we do not
  discuss it here because the results that we are going to use in the next
  section can be tested directly and easily.

\section{Violation of Equalities and  Inequalities \label{sec_03}}
  Using the numerical method mentioned in the previous section we have found
  two expressions in $22\!\otimes\!22$ case which are complement of each other.
  These are
  \be
     |P_{11}^{10} - P_{12}^{11} + P_{21}^{11} + P_{22}^{01}|=
     q_{1212}^{0001}+q_{1212}^{0011}+q_{1212}^{0110}+q_{1212}^{0111}+
     q_{1212}^{1000}+q_{1212}^{1001}+q_{1212}^{1100}+q_{1212}^{1110},
       \label{eq_021}
  \ee
  and
  \be
     |-P_{11}^{01} + P_{12}^{00} + P_{21}^{01} + P_{22}^{11}|=
     q_{1212}^{0000}+q_{1212}^{0010}+q_{1212}^{0100}+q_{1212}^{0101}+
     q_{1212}^{1010}+q_{1212}^{1011}+q_{1212}^{1101}+q_{1212}^{1111}.
       \label{eq_022}
  \ee
  Please note that all $q's$ are non-negative numbers. Adding these two
  equations and then using eq.~(\ref{eq_001}), would give
  \be
     |P_{11}^{10} - P_{12}^{11} + P_{21}^{11} + P_{22}^{01}| +
     |-P_{11}^{01} + P_{12}^{00} + P_{21}^{01} + P_{22}^{11}| = 1.
       \label{eq_023}
  \ee
  See Appendix~\ref{app_01} for a direct proof of eqs.~(\ref{eq_021}),
  (\ref{eq_022}) and the above equality.
  (Though the equality~(\ref{eq_023}) can also be derived analytically from
  eqs.~(\ref{eq_004} -- \ref{eq_014}) easily.)
  Of course, due to symmetry, there are other
  equations of this type as well which we do not mention here for brevity. 

  Now we use the same experiment used by CH \cite{CH_74} to show that
  {\em equality} (\ref{eq_023}) is violated
  by quantum theory. Consider a two photon system in the state
  \be
    |\Psi_0\rangle =\frac{1}{\sqrt{2}}
      \left[
        \left(\begin{array}{c} 1 \\ 0 \\ 0 \end{array}\right)\otimes
        \left(\begin{array}{c} 1 \\ 0 \\ 0 \end{array}\right) +
        \left(\begin{array}{c} 0 \\ 1 \\ 0 \end{array}\right)\otimes
        \left(\begin{array}{c} 0 \\ 1 \\ 0 \end{array}\right)
      \right]
    \label{eq_024}
  \ee
  where one photon moves to the left in the $+z$ direction and the other moves
  to the right in $-z$ direction. The projection of $|\Psi_0\rangle$ on
  directions $\mathbf{u}(\theta)=\cos\theta \mathbf{i} + \sin\theta \mathbf{j}
  + 0\mathbf{k}$ (for the left photon) and $\mathbf{v}(\phi)=\cos\phi
  \mathbf{i} + \sin\phi \mathbf{j} + 0\mathbf{k}$ (for the right photon) is:
  \be
    |\Psi(\theta,\phi)\rangle =
    \left(
    \begin{array}{ccc}
      \cos^2(\theta)           & \cos(\theta)\sin(\theta) & 0 \\
      \cos(\theta)\sin(\theta) & \sin^2(\theta)           & 0 \\
      0                        & 0                        & 0
    \end{array}
    \right)\otimes
    \left(
    \begin{array}{ccc}
      \cos^2(\phi)           & \cos(\phi)\sin(\phi) & 0 \\
      \cos(\phi)\sin(\phi)   & \sin^2(\phi)         & 0 \\
      0                      & 0                    & 0
    \end{array}
    \right)
    |\Psi_0\rangle. \nonumber
  \ee
  So
  \be
    |\Psi(\theta,\phi)\rangle = \frac{1}{\sqrt{2}}\cos(\theta-\phi)
    \left(\begin{array}{c} \cos\theta \\ \sin\theta \\ 0
          \end{array}\right)\otimes
    \left(\begin{array}{c} \cos\phi \\ \sin\phi \\ 0 \end{array}\right).
    \label{eq_025}
  \ee
  If for example the left photon is not detected in the $\mathbf{u}(\theta)$
  direction then it must be detected in the direction
  $\mathbf{u}(\theta+\pi/2)=-\sin\theta \mathbf{i} + \cos\theta \mathbf{j} +
  0\mathbf{k}$ which is perpendicular to $\mathbf{u}(\theta)$. So defining
  $\gamma=\theta+\pi/2$ and $\eta=\phi+\pi/2$, for directions perpendicular to
  $\mathbf{u}$ and $\mathbf{v}$ respectively, we obtain:
  \be
    |\Psi(\gamma,\eta)\rangle = \frac{1}{\sqrt{2}}\cos(\theta-\phi)
    \left(\begin{array}{c} -\sin\theta \\ \cos\theta \\ 0
          \end{array}\right)\otimes
    \left(\begin{array}{c} -\sin\phi \\ \cos\phi \\ 0 \end{array}\right),
    \label{eq_026}
  \ee
  \be
    |\Psi(\theta,\eta)\rangle = \frac{1}{\sqrt{2}}\sin(\phi-\theta)
    \left(\begin{array}{c} -\sin\theta \\ \cos\theta \\ 0
          \end{array}\right)\otimes
    \left(\begin{array}{c}  \cos\phi \\ \sin\phi \\ 0 \end{array}\right),
    \label{eq_027}
  \ee
  \be
    |\Psi(\gamma,\phi)\rangle = \frac{1}{\sqrt{2}}\sin(\theta-\phi)
    \left(\begin{array}{c} \cos\theta \\ \sin\theta \\ 0
          \end{array}\right)\otimes
    \left(\begin{array}{c} -\sin\phi \\ \cos\phi \\ 0 \end{array}\right).
    \label{eq_028} 
  \ee
  Denoting the photon detection in a particular direction with~1 and the
  non-detection with~0 the joint probabilities would be:
  \be
    \begin{array}{llll}
    P_{11}^{11} = \frac{1}{2}\cos^2(\theta - \phi); &
    P_{11}^{10} = \frac{1}{2}\sin^2(\phi - \theta); &
    P_{11}^{01} = \frac{1}{2}\sin^2(\theta - \phi); &
    P_{11}^{00} = \frac{1}{2}\cos^2(\theta - \phi); \nonumber \\

    P_{12}^{11} = \frac{1}{2}\cos^2(\theta - \phi'); &
    P_{12}^{10} = \frac{1}{2}\sin^2(\phi' - \theta); &
    P_{12}^{01} = \frac{1}{2}\sin^2(\theta - \phi'); &
    P_{12}^{00} = \frac{1}{2}\cos^2(\theta - \phi'); \nonumber \\

    P_{21}^{11} = \frac{1}{2}\cos^2(\theta' - \phi); &
    P_{21}^{10} = \frac{1}{2}\sin^2(\phi - \theta'); &
    P_{21}^{01} = \frac{1}{2}\sin^2(\theta' - \phi); &
    P_{21}^{00} = \frac{1}{2}\cos^2(\theta' - \phi); \nonumber \\

    P_{22}^{11} = \frac{1}{2}\cos^2(\theta' - \phi'); &
    P_{22}^{10} = \frac{1}{2}\sin^2(\phi' - \theta'); &
    P_{22}^{01} = \frac{1}{2}\sin^2(\theta' - \phi'); &
    P_{22}^{00} = \frac{1}{2}\cos^2(\theta' - \phi'). \nonumber \\

    \end{array}
  \ee
  If we represent $\theta - \phi$, $\phi' - \theta$ and $\phi - \theta'$ by
  $x$, $y$ and $z$ respectively then $\theta' - \phi'$, denoted by $w$, would
  be $-(x+y+z)$.

  For $x=247.46^\circ$, $y=67.49^\circ$ and $z=157.50^\circ$ the value of the
  second term on the left hand side of eq.~(\ref{eq_023}), predicted by
  quantum theory, is
  $|-1/2\sin^2x + 1/2\cos^2y + 1/2\sin^2z + 1/2\cos^2w|=0.207$
  and if quantum theory is local, according to eq.~(\ref{eq_023}), for the
  first term we must have
  \be
    |P_{11}^{10} - P_{12}^{11} + P_{21}^{11} + P_{22}^{01}| = 0.793.
    \label{eq_029}
  \ee

  However, the value of the left hand side of the above equality
  predicted by quantum theory is
  $|+1/2\sin^2x - 1/2\cos^2y + 1/2\cos^2z + 1/2\sin^2w|=1.207$.
  Clearly equality~(\ref{eq_029}) is violated by quantum theory by a
  factor of $1.522$ which exceeds that of CH results by $0.108$.
  
  In the presence of white noise the density matrix is
  \be
    \rho=\gamma\rho_{noise}+(1-\gamma)\rho_{QM}
    \label{eq_030}
  \ee
  where $\rho_{noise}=(1/4)\times\openone$ and
  $\rho_{QM}=|\Psi\rangle\langle\Psi|$. Here $\gamma$ is the Bell expression's
  tolerance of white noise {\it i.e.} the maximum fraction of white noise
  admixture for which a Bell expression stops being violated. The joint
  probability for the above state is
  \be
    {\cal P}_{ab}^{ij}=\frac{\gamma}{4}+(1-\gamma)P_{ab}^{ij}
    \label{eq_031},
  \ee
  and it is easily seen that the tolerance of equality~(\ref{eq_029}) is
  $0.586$ fraction of white noise which is twice that of CH inequality.
  This is due to the fact that the tolerance is very sensitive to upper bound
  which is 1 in CH inequality and 0.793 in our {\em equality}.

  Also using the method mentioned in section~\ref{sec_03} we have
  obtained three inequalities which correspond to different
  values of $c$ and $d$ in eq.~(\ref{eq_020}) which can be tested directly too.
  One of them, for $c=1$ and $d=0$, is Clauser-Horne inequality, which reads
  \be
    0 \le + P_{11}^{10} + P_{12}^{00} + P_{21}^{01} - P_{22}^{00} \le 1.
         \label{eq_032}
  \ee
  The other, for $c=1$ and $d=1$, is
  \be
    -1 \le + 2P_{11}^{10} + P_{12}^{00} - P_{12}^{10}
           - P_{21}^{00} + P_{21}^{01} - P_{22}^{00} - P_{22}^{11} \le 1.
         \label{eq_033}
  \ee
  Finally, for $c=2$ and $d=1$, we found
  \be
    -1 \le 
      - P_{11}^{00} + P_{11}^{01} + P_{11}^{10}
      + P_{12}^{00} - P_{12}^{10} + P_{12}^{11}
      + P_{21}^{01} - 2P_{21}^{11}
      - P_{22}^{00} + 2P_{22}^{10}
    \le 2.
    \label{eq_034}
  \ee

  A direct proof of this inequality is shown in Appendix~\ref{app_02}.
  In terms of $x$, $y$, $z$ and $w$, the inequalities~(\ref{eq_032}), (\ref{eq_033}) and
  (\ref{eq_034}) would become
  \be
    0 \le \frac{1}{2}( \sin^2 x + \cos^2 y  + \sin^2 z - \cos^2 w ) \le 1
    \label{eq_035},
  \ee
  \be
    -1 \le \sin^2 x + \frac{1}{2}\cos2y -
           \frac{1}{2}\cos 2z  - \cos^2 w ) \le 1
    \label{eq_036},
  \ee  
  \be
    -1 \le \frac{1}{2}+\frac{3}{2}(\sin^2 x -\sin^2 y -\cos^2 z +\sin^2 w)
    \le 2
    \label{eq_037},
  \ee
  respectively.

  The value of inequalities (\ref{eq_035}), (\ref{eq_036}) and (\ref{eq_037})
  for $x=-67.50^\circ(-3\pi/8)$, $y=202.50^\circ(9\pi/8)$ and
  $z=-67.50^\circ(-3\pi/8)$ are $+1.207$, $+1.414$ and
  $+2.621$ respectively; but for $x=22.50^\circ(\pi/8)$,
  $y=-67.50^\circ(-3\pi/8)$ and $z=22.50^\circ(\pi/8)$ these are
  $-0.207$, $-1.414$ and $-1.621$ respectively. While the difference between
  upper bound and lower bound for Clauser-Horne inequality is $1.414$, in
  agreement with the result obtained in \cite{CH_74}, in eq.~(\ref{eq_033}) the
  lower bound itself is violated by a factor of $1.414$ and the lower bound of
  eq.~(\ref{eq_034}) is violated by a factor of $1.621$ which are greater
  than that obtained by others for $22\!\otimes\!22$ case (see {\it e.g.}
  \cite{CHSH_69} and
  \cite{CH_74}). Please note that the difference between the upper bound and
  the lower bound in the two latter cases are 2.818 and 4.242 respectively and
  that the range of violation for the upper bound for
  inequalities~(\ref{eq_032}), (\ref{eq_033}) and (\ref{eq_034}) are
  $0.207$, $0.414$ and $0.621$ respectively.
  These are the maximum values that we could reach and despite the greater
  range of violation, inequalities~(\ref{eq_032}), (\ref{eq_033}) and
  (\ref{eq_034}) tolerate $0.293$ of white noise which is the same as
  CH inequality.

\section{Conclusion} \label{sec_04}
  In this paper we have used a numerical method to derive all possible Bell
  type inequalities in a two party experiment namely $mm'\!\otimes\!nn'$. The
  dimension of joint probabilities space obtained by this method is not in
  agreement with the results obtained previously in the literature. However we
  confirmed our numerical results using an analytical method. This was done by
  adopting the assumption of CH, {\it i.e.} eq.~(\ref{eq_008}). And on this basis
  we showed that {\it normalization} of joint probabilities in a run of
  experiment and {\it no-signaling} in local theories are implied by
  eq.~(\ref{eq_001}) which is statistically trivial.  It is an open question
  that if any assumption other than eq.~(\ref{eq_008}) could leads to these
  correct results.

  We showed that besides inequalities there are equalities that can be used to
  test local theories. In $22\!\otimes\!22$ case the equality derived here is
  violated by a factor of $1.522$ tolerating 0.586 fraction of white noise
  admixture which is twice that of the previous restults in the literature.
  However as in higher dimensions there are inequalities which are violated by
  a stronger factor (see~\cite{MOV_H_98}), we expect there may exist equalities
  in higher dimensions as well which are violated by a stronger factor.

  Among many possible Bell type inequalities that we derived, we could only
  find three types in the form of eqs.~(\ref{eq_032}), (\ref{eq_033}) and
  (\ref{eq_034}) which are violated with the same settings used in
  \cite{CHSH_69} and \cite{CH_74}. However there are some other settings too
  for which these inequalities are violated and for all of them the inequality
  in the form of eq.(~\ref{eq_034}) is violated with a stronger factor of 1.621
  tolerating $0.293$ fraction of white noise admixture.
  
\section{Acknowledgment} \label{sec_05}
  This paper is dedicated to late Professor Euan James Squires, my Ph.D.
  supervisor and the author of 'The Mystery of the Quantum World' to whom I owe
  much more than words can describe.

\newpage
\appendix
\section{Direct Proof of Equality~(\ref{eq_023}) \label{app_01}}

  For $22\!\otimes\!22$ case, using the definition of joint probability $P$
  in terms of $q$'s, {\it i.e.} eq.~(\ref{eq_002}), there are $16$ possible
  $P$'s which are:
  \begin{displaymath}
     \begin{array}{lllllllll}
        P_{11}^{00} & = & q_{1212}^{0000} & + & q_{1212}^{0001} & + & q_{1212}^{0100} & + & q_{1212}^{0101} \\
        P_{11}^{01} & = & q_{1212}^{0010} & + & q_{1212}^{0011} & + & q_{1212}^{0110} & + & q_{1212}^{0111} \\
        P_{11}^{10} & = & q_{1212}^{1000} & + & q_{1212}^{1001} & + & q_{1212}^{1100} & + & q_{1212}^{1101} \\
        P_{11}^{11} & = & q_{1212}^{1010} & + & q_{1212}^{1011} & + & q_{1212}^{1110} & + & q_{1212}^{1111} \\
		      	  
        P_{12}^{00} & = & q_{1212}^{0000} & + & q_{1212}^{0010} & + & q_{1212}^{0100} & + & q_{1212}^{0110} \\
        P_{12}^{01} & = & q_{1212}^{0001} & + & q_{1212}^{0011} & + & q_{1212}^{0101} & + & q_{1212}^{0111} \\
        P_{12}^{10} & = & q_{1212}^{1000} & + & q_{1212}^{1010} & + & q_{1212}^{1100} & + & q_{1212}^{1110} \\
        P_{12}^{11} & = & q_{1212}^{1001} & + & q_{1212}^{1011} & + & q_{1212}^{1101} & + & q_{1212}^{1111} \\
		      	  
        P_{21}^{00} & = & q_{1212}^{0000} & + & q_{1212}^{0001} & + & q_{1212}^{1000} & + & q_{1212}^{1001} \\
        P_{21}^{01} & = & q_{1212}^{0010} & + & q_{1212}^{0011} & + & q_{1212}^{1010} & + & q_{1212}^{1011} \\
        P_{21}^{10} & = & q_{1212}^{0100} & + & q_{1212}^{0101} & + & q_{1212}^{1100} & + & q_{1212}^{1101} \\
        P_{21}^{11} & = & q_{1212}^{0110} & + & q_{1212}^{0111} & + & q_{1212}^{1110} & + & q_{1212}^{1111} \\
		      	  
        P_{22}^{00} & = & q_{1212}^{0000} & + & q_{1212}^{0010} & + & q_{1212}^{1000} & + & q_{1212}^{1010} \\
        P_{22}^{01} & = & q_{1212}^{0001} & + & q_{1212}^{0011} & + & q_{1212}^{1001} & + & q_{1212}^{1011} \\
        P_{22}^{10} & = & q_{1212}^{0100} & + & q_{1212}^{0110} & + & q_{1212}^{1100} & + & q_{1212}^{1110} \\
        P_{22}^{11} & = & q_{1212}^{0101} & + & q_{1212}^{0111} & + & q_{1212}^{1101} & + & q_{1212}^{1111}
     \end{array}
  \end{displaymath}

  Let's write eq.~(\ref{eq_023}) as
  \be
     E = |E_1| + |E_2|
  \ee
  where $E_1$ and $E_2$ are
  \be
     E_1 = P_{11}^{10} - P_{12}^{11} + P_{21}^{11} + P_{22}^{01}
  \ee
  and
  \be
     E_2 = - P_{11}^{01} + P_{12}^{00} + P_{21}^{01} + P_{22}^{11}.
  \ee
  Replacing $P$'s in terms of $q$'s we get
  \bean
       E_1 & = &
           + q_{1212}^{1000} + q_{1212}^{1001} + q_{1212}^{1100} + q_{1212}^{1101}
                \\
       & &
           - q_{1212}^{1001} - q_{1212}^{1011} - q_{1212}^{1101} - q_{1212}^{1111}
                \\
       & &
           + q_{1212}^{0110} + q_{1212}^{0111} + q_{1212}^{1110} + q_{1212}^{1111}
                \\
       & &
           + q_{1212}^{0001} + q_{1212}^{0011} + q_{1212}^{1001} + q_{1212}^{1011}
                \\
       &  = &
             q_{1212}^{0001} + q_{1212}^{0011} + q_{1212}^{0110} + q_{1212}^{0111}
           + q_{1212}^{1000}+q_{1212}^{1001}+q_{1212}^{1100}+q_{1212}^{1110},
  \eean
  \bean
       E_2 & = &
           - q_{1212}^{0010} - q_{1212}^{0011} - q_{1212}^{0110} - q_{1212}^{0111}
                \\
       & &
           + q_{1212}^{0000} + q_{1212}^{0010} + q_{1212}^{0100} + q_{1212}^{0110}
                \\
       & &
           + q_{1212}^{0010} + q_{1212}^{0011} + q_{1212}^{1010} + q_{1212}^{1011}
                \\
       & &
           + q_{1212}^{0101} + q_{1212}^{0111} + q_{1212}^{1101} + q_{1212}^{1111}
                \\
       & = &
             q_{1212}^{0000} + q_{1212}^{0010} + q_{1212}^{0100} + q_{1212}^{0101}
           + q_{1212}^{1010} + q_{1212}^{1011} + q_{1212}^{1101} + q_{1212}^{1111}.
  \end{eqnarray*}
  However as $q$'s are all non-negative numbers, $E_1 = |E_1|$ and
  $E_2 = |E_2|$. So for local theories we may write
  \bean
           E & = & |E_1| + |E_2|
                \\
       & = &
           + q_{1212}^{0001} + q_{1212}^{0011} + q_{1212}^{0110} + q_{1212}^{0111}
           + q_{1212}^{1000}+q_{1212}^{1001}+q_{1212}^{1100}+q_{1212}^{1110}
                \\
       & &
           + q_{1212}^{0000} + q_{1212}^{0010} + q_{1212}^{0100} + q_{1212}^{0101}
           + q_{1212}^{1010} + q_{1212}^{1011} + q_{1212}^{1101} + q_{1212}^{1111}
                \\
       & = &
             1
  \end{eqnarray*}
  where the last equality is implied by eq.~(\ref{eq_001}) and this
  justifies eq.~(\ref{eq_023}).

\section{Direct Proof of Inequality~(\ref{eq_034}) \label{app_02}}
  With a similar procedure used in Appendix~\ref{app_01}, for
  inequality~(\ref{eq_034}) we have:
  \bean
           J & = & - P_{11}^{00} + P_{11}^{01} + P_{11}^{10}
           + P_{12}^{00} - P_{12}^{10} + P_{12}^{11}
           + P_{21}^{01} - 2P_{21}^{11}
           - P_{22}^{00} + 2P_{22}^{10}
              \\
      & = &
           - q_{1212}^{0000} - q_{1212}^{0001} - q_{1212}^{0100} - q_{1212}^{0101}
                \\
      & &
           + q_{1212}^{0010} + q_{1212}^{0011} + q_{1212}^{0110} + q_{1212}^{0111}
                \\
      & &
           + q_{1212}^{1000} + q_{1212}^{1001} + q_{1212}^{1100} + q_{1212}^{1101}
                \\
      & &
           + q_{1212}^{0000} + q_{1212}^{0010} + q_{1212}^{0100} + q_{1212}^{0110}
                \\
      & &
           - q_{1212}^{1000} - q_{1212}^{1010} - q_{1212}^{1100} - q_{1212}^{1110}
                \\
      & &
           + q_{1212}^{1001} + q_{1212}^{1011} + q_{1212}^{1101} + q_{1212}^{1111}
                \\
      & &
           + q_{1212}^{0010} + q_{1212}^{0011} + q_{1212}^{1010} + q_{1212}^{1011}
                \\
      & &
      - 2 (  q_{1212}^{0110} + q_{1212}^{0111} + q_{1212}^{1110} + q_{1212}^{1111} )
                \\
      & &
           - q_{1212}^{0000} - q_{1212}^{0010} - q_{1212}^{1000} - q_{1212}^{1010}
                \\
      & &
      + 2 (  q_{1212}^{0100} + q_{1212}^{0110} + q_{1212}^{1100} + q_{1212}^{1110} )
                \\
      & = &
         - ( q_{1212}^{0000} + q_{1212}^{0001} + q_{1212}^{0101} + q_{1212}^{0111}
           + q_{1212}^{1000} + q_{1212}^{1010} + q_{1212}^{1110} + q_{1212}^{1111}
           )
                \\
      & &
       + 2 ( q_{1212}^{0010} + q_{1212}^{0011} + q_{1212}^{0100} + q_{1212}^{0110}
           + q_{1212}^{1001} + q_{1212}^{1011} + q_{1212}^{1100} + q_{1212}^{1101}
           )
  \eean
  Since according to eq.~(\ref{eq_001}) each of the parentheses on the right
  hand-side of the above equation is
  less than 1 and greater than zero,
  it is easily seen that $-1 \le J \le2$ which is inequality~(\ref{eq_034}).

\end{document}